# Mode-locked Lasers Applied to Deflecting a Near Earth Object on Collision Course with Earth

**Richard Fork, Spencer Cole, Luke Burgess, and Grant Bergstue**

*Electrical and Computer Engineering Department, University of Alabama Huntsville,*
*301 Sparkman Drive, Huntsville, AL 35899*
Corresponding author: *forkr@uah.edu*

We consider synchronized trains of sub-picosecond pulses generated by mode-locked lasers applied to deflection of near Earth objects (NEO) on collision course with Earth. Our method is designed to avoid a predicted collision of the NEO with Earth by at least the diameter of Earth. We estimate deflecting a 10,000 MT NEO, such as the asteroid which struck Earth near Chelyabinsk, Russia to be feasible within several months using average power in the ten kilowatt range. We see this deflection method as scalable to larger NEO to a degree not possible using continuous laser systems.

Referring to near Earth objects (NEO) discovered to be on a collision course with Earth, the White House Office of Science and Technology charged NASA on October 15, 2010 with "implementing a deflection campaign, in consultation with international bodies, should one be necessary" [1]. We describe here a means of using energetic subpicosecond optical pulses to deflect NEO. Mode-locked laser systems [2, 3] have evolved over nearly a half century. We consider ultrashort optical pulses [4, 5] applied to generating ablative propulsive thrust [6] so as to deflect NEO. We consider pulse energies, e.g., 40 mJ, and duration, e.g. < 1 picosecond, within current capabilities [4,5]. Pulses of this peak energy, and substantially larger peak energy, can be used freely in the vacuum of space. These pulses, however, cannot be usefully transmitted over distances of interest in the atmosphere of Earth [7].

The force $F = 2P/v_{ej}$ [8] exerted by using the material of the NEO itself as propellant ejected at a velocity $v_{ej}$ of ~ 10,000 m/sec is 8 MN. This thrust, *during the brief time it occurs*, exceeds the force generated by the main engine of the, recently retired, Space Shuttle, 2.3 MN [9]. The large force exerted during the ablation event and the absence of a need to provide propellant makes this strategy we outline here valuable for use in space. This is especially true in the deeper regions of space where delivery of propellant can become prohibitively expensive. We see the principal current challenges as scaling this strategy in average power by increasing the repetition rate of the optical pulses and optimizing the delivery of the deflection impulses to the NEO. We see advances in synchronization of trains of pulses generated by modelocked lasers, e.g., the 2005 Nobel Prize winning work in physics [10, 11], as relevant. We suggest the usual constraint on scaling laser power posed by heat generated in the laser gain media can be minimized. The lasers can be physically separated so as to facilitate thermal management, but still precisely synchronized so multiple physically well separated lasers can be used in parallel.

We consider an example of deflecting impulses applied by three or more pulses delivered to multiple locations on the NEO with sub-picosecond temporal resolution. The cooperative group of three or more pulses can deliver an impulse designed so the net deflective impulse is optimally directed through the center of mass of the NEO, while the correlated, but individual, ablative events are still widely distributed on the NEO, Fig. 1. Three, or more, points used in combination with adjustment of the magnitudes of each force enable not only slowing, but also steering, of the NEO.

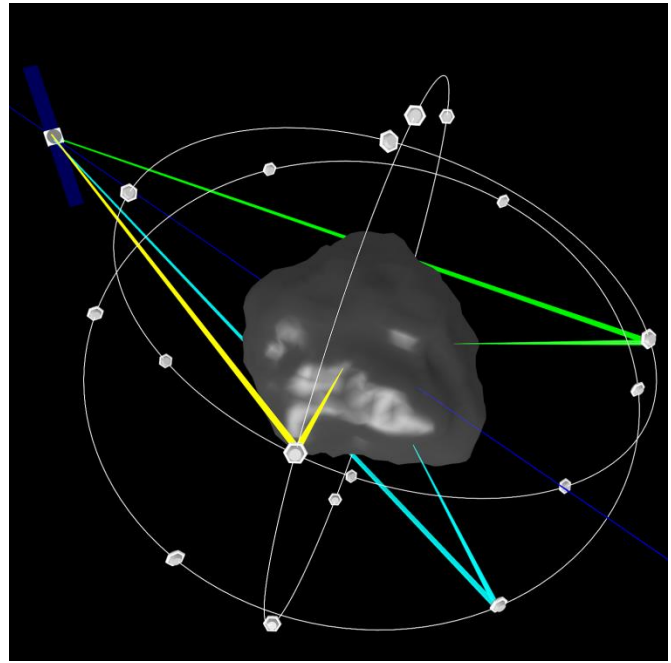

Fig.1. Three trains of synchronized sub‐picosecond optical pulses (light blue, green, and yellow Gaussian $TEM_{00}$ modes) directed by microspacecraft (white hexagonal boundaries) exert propulsive thrust slowing a NEO.

Colors are used in Fig. 1 for illustrational purposes only. The microspacecraft in this example are organized [12] in 3 orbital planes orthogonal to each another. Each orbital plane contains six micro-satellites which for one sixth of their orbit are positioned so as reflect and focus multiple lowest order

Gaussian beams onto the leading surface of the NEO. Each train of pulses transmitted from the spacecraft is then redirected to, and focused on, the surface of the NEO using a turning angle less than 15 degrees and f-number greater than 200 to limit astigmatic distortion.

We propose, e.g., a system of lasers and microspacecraft that: (1) Obtain and continually update a sub-millimeter resolution 3-D map of the NEO surface; (2) use the information in that map to apply the laser pulses, in groups of three or more, so as to cause optimal net deflecting force on the NEO. The microspacecraft configuration illustrated in Fig. 1 is designed to enable: (1) creation and continuous maintenance of the high resolution 3-D map; and (2) delivery of the optical pulses so as to apply the forces as illustrated in Fig. 2 and described in Eq. 1. The pulses can, e.g., be focused to areas < 1 mm diameter at pulse repetition rates up to hundreds of kHz for a given location.

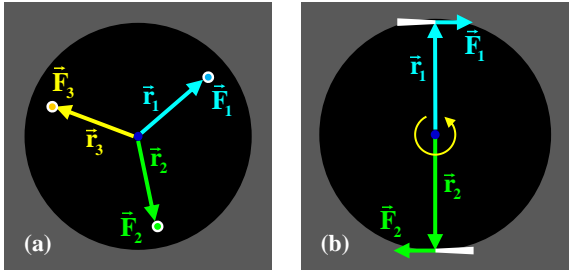

Fig. 2. Forces $\vec{F}_n$, position vectors $\vec{r}_n$, and center of mass (dark blue) of the NEO.

The plumes of ejecta are shown as white dots in (a) and as white extended trapezoidal shapes in (b). The plumes are normal to the plane of the figure in (a) and within the plane of the figure tangential to the radial vectors $\vec{r}_n$ in (b). The ablated material is ejected approximately normal to each illuminated surface region. We do not specify the means of finding or creating favorable surface locations, but note that the ablative capability of the ultrashort duration creates options for shaping such surfaces. In Fig. 2 (a) three forces $\vec{F}_n$ are applied simultaneously within sub-picosecond resolution. The positions and orientations of the surface normal where these forces are applied in Fig. 2 (a) are chosen to maximize the rate of slowing of the NEO while avoiding coupling energy into unwanted rotation of the NEO. Small variations in the locations and surface normal of the ablation events can be used for fine steering of the deflection process. Referring to Fig. 2 (a) the condition for optimum application of the slowing force $\vec{F}$ can be written as

$$\vec{F} = \sum_{n=1}^{3} \vec{F}_n, \quad \vec{F}_n \cong \hat{z}|F_n|, \quad \sum_{n=1}^{3} \vec{r}_n \times \vec{F}_n \cong 0 \quad (1)$$

The NEO in Fig. 1 is assumed comparable in dimensions to the asteroid which struck Earth near Chelyabinsk, e.g., 20 m diameter. The directions of arrival of the optical pulses at the NEO surface are relatively unimportant because of the small photon momentum. For a laser wavelength of 1 micron, a microspacecraft orbital radius of ~ 30 m, and a ~10 m radius NEO the diameter of the optical aperture on the microsatellites needs to be > 10 cm. Pulses to the same location on the NEO need to be separated in time by > 3 microseconds to avoid loss due to the transient absorbing plume of ejecta [13]. We see this use of temporally resolved pulses as not only valuable, but essential to avoiding absorptive loss in the ejecta associated with ablation. We see this strategy as both adequate to enable use of large average power and as also most likely to rule out most strategies based on simplistic vaporization of NEO.

Another task, de-spinning (stopping the rotational motion) of a NEO appears addressable. Favorable conditions are as described in Eq. (2) and illustrated in Fig. 2 (b). Referring to Fig. 2 (b) the torque on the NEO is

$$\vec{T}_R = \sum_{n=1}^{2} \vec{r}_n \times \vec{F}_n = \hat{n}|r_1||F_1||r_2||F_2| \quad (2)$$

We use the unit vector $\hat{n}$ to specify the direction of the torque exerted on the NEO.

We calculate the time $\Delta t$ and power P required for a given change in arrival time of a NEO at the location where a collision between the NEO and Earth has been predicted to occur. We assume near unit ablative efficiency [6]. Here $v_{ej}$ is the velocity of the material ejected from the NEO. The time $\Delta t$ required to slow a NEO of mass M so it still has distance $D_E$ (diameter of Earth) to travel to the location of Earth at the predicted collision time is

$$\Delta t = \sqrt{D_E M v_{ej} / P} \quad (3)$$

We plot in Fig. 3 the time to achieve deflection $D_E$ vs. optical power for several NEO. We do not include the cost in energy to remove waste heat or perform tasks other than the deflection. For Fig. 3 we assume an ejection velocity of ablated material $v_{ej}$ of $10^4$ m/s [6].

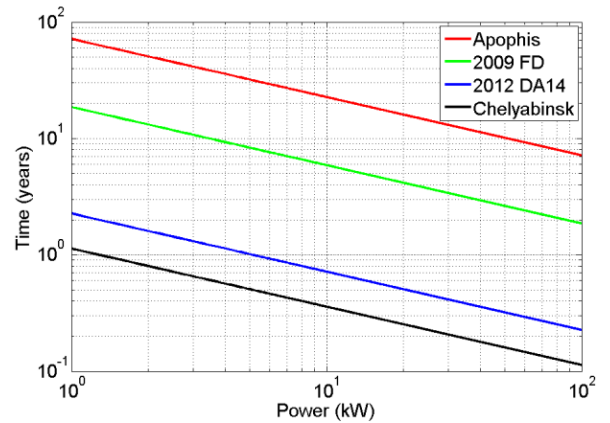

Fig. 3. Time and power required to deflect a NEO of mass M so as to avoid a collision with Earth by distance $D_E$.

The examples used in Fig. 3 are: (1) the Chelyabinsk meteor estimated to have mass 10,000 MT and diameter ~17-20 m, which struck Earth 15 February 2013 [14]; (2) the asteroid 2012 DA14 of mass 40,000 MT and dimensions, 20m x 40m, which made a record close approach to Earth, 27.7 Mm, in 2013 [15]; (3) the asteroid 2009 FD, an Apollo class asteroid rated -1.80 on the Palermo Scale, placing it relatively high on the Sentry Risk Table, of mass $2.7 \times 10^9$ kg and diameter 130 m, which is due back near Earth in 2185 [16]; and (4) the asteroid Apophis of mass $4 \times 10^{10}$ kg and diameter 325 m [17].

Our calculation implies smaller NEO can be deflected in useful periods of time using multiple synchronized laser systems generating power in ranges already demonstrated [4,5]. The scale of the threat posed by NEO to life on Earth, however, becomes more evident when we consider some of the larger long period comets. Hale-Bopp, e.g., has a mass of $10^{16}$ kg [18] which is nine orders of magnitude larger than the Chelyabinsk asteroid. Also comets arrive from virtually any direction and may be discovered with only a fraction of a year remaining before arrival near Earth.

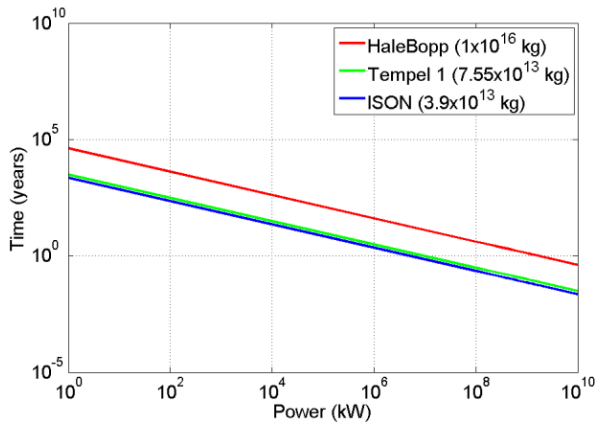

Fig. 4. Minimum time vs. power for deflecting larger comets. assuming material ejected at $10^4$ m/s.

We plot time vs. power for deflection of some NEO assuming material ejected at $10^4$ m/s in Fig. 4. The time and power requirements are substantial.. Comets are typically dust and ice with a thin exterior layer of material that has been largely depleted of water [19]. This could offer some options not available with typical asteroids, such as first de-spinning the comet and then ejecting material at velocities closer to the escape velocity. We calculate a minimum time for stopping rotation (de-spinning) of a NEO of mass M, radius R, and period $\tau$ as

$$\Delta t_{ds} = (2\pi / 5P\tau) M R v_{ej} \qquad (4)$$

This is an interesting area of investigation, but will presumably require much more thorough exploration of, and ability to work in deeper regions of space surrounding our Sun, such as the Kuiper Belt. Using an ejection velocity of $10^4$ m/s we obtain a plot of time to de-spin a comet vs. power in Fig.5.

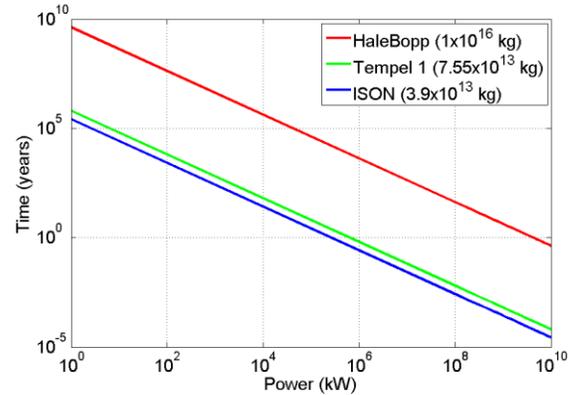

Fig. 5. Minimum time vs. power for stopping rotation of a NEO assuming material ejected at $10^4$ m/s.

Also the synchronization of multiple modelocked lasers, as facilitated, e.g., by frequency comb modelocking [10, 11] further facilitates this multiplexing. This strategy does, however, strongly favor the close proximity of the spacecraft and NEO, such as, illustrated in Fig. 1. We recommend the grouping of the source lasers in a group having a minimum number of three precisely timed and synchronized modelocked lasers so as to maintain the timing and allow the strategies illustrated in Fig. 2. That is we write the power requirement as

$$P = D_E M v_{ej} / (\Delta t)^2 = N_{LG} P_{LG} \qquad (6)$$

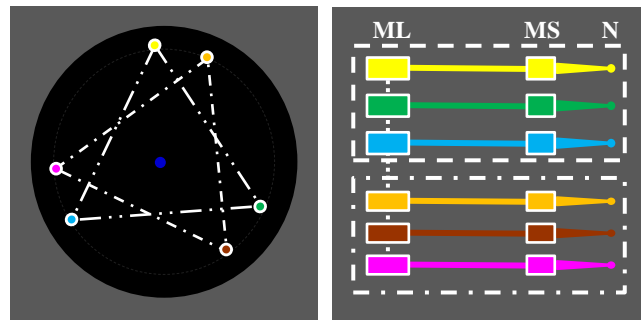

Fig.6. Strategy for delivering deflection energy using groups of three pulses each. Pulses within a given group (defined by segmented lines), and only within a given group, are synchronized with each other with sub-picosecond precision (ML indicates the mode-locked laser system, MS a microspacecraft, and N the NEO).

Here $N_{LG}$ denotes the number of groups of laser pulses and $P_{LG}$ the power delivered by a single group, of pulses. The point is the sub-picosecond timing precision is required for the relative timing of pulses within the small group of pulses that apply the deflecting power, as illustrated in Fig. 2.

We see the overall task as primarily a task in information acquisition, management and continuous evaluation of the degree of success in application of that information. The proximity of the spacecraft to the NEO, typically a few kilometers, and the extended time used for deflection offer opportunities for continuous monitoring and further optimization of the deflection process. We see the close spatial relationship of the NEO and the system for applying the deflecting thrust as essential for this application. The long period for application of the deflecting impulse appears to require early identification and characterization of NEO on collision course with Earth.

We observe incidentally that the station keeping energy of the co-orbiting spacecraft and NEO can be reduced close to zero, for some conditions, by balancing the radiation reaction force of the laser emission on the spacecraft with the gravitational attraction between the spacecraft and the NEO. This requires a distance between the spacecraft and NEO of

$$d = \sqrt{cGM_sM/P} \qquad (7)$$

Here G is the gravitational constant; c is the velocity of light and $M_s$ is the mass of the spacecraft. For an asteroid of mass M = 10,000 MT, a spacecraft of 10 MT, and average optical power of 10 kW, this distance d is, e.g., ~ 0.5 km. This is essentially an "Archimedes Lever" strategy that treats the spacecraft, micro-spacecraft and NEO as a single unit. This configuration might be of interest for other applications.

The general concept of using ultrashort pulses to deflect asteroids was presented by two of the authors at an earlier NASA conference on this topic [20]. An earlier description of some of the concepts considered here is available as a point of view paper in IEEE Proceedings [21]. We thank Pat Reardon, Ken Pitalo, and Dave Pollock for useful discussions and the University of Alabama in Huntsville for support of this work. We also greatly appreciate advice and support from colleagues at NASA Marshall Space Flight Center, NASA Glenn Research Center and others active in space related activities.

A unique feature of pulsed, and consequently of modelocked lasers, that distinguish them from continuous lasers is that the laser systems can be time domain multiplexed as a means of delivering power in ways differing from continuous lasers and advantageous for deflection of NEO. The interleaving of pulses only requires a timing precision of the order of the pulse duration rather than of the carrier frequency.